# Bitcoin and quantum computing

Louis Tessler[2,6] and Tim Byrnes[1,2,3,4,5]

*1 State Key Laboratory of Precision Spectroscopy, School of Physical and Material Sciences,*

*East China Normal University, Shanghai 200062, China*

*2 New York University Shanghai, 1555 Century Ave, Pudong, Shanghai 200122, China*

*3 NYU-ECNU Institute of Physics at NYU Shanghai,*

*3663 Zhongshan Road North, Shanghai 200062, China*

*4 National Institute of Informatics, 2-1-2 Hitotsubashi, Chiyoda-ku, Tokyo 101-8430, Japan*

*5 Department of Physics, New York University, New York, NY 10003, USA*

*6 CEMS, RIKEN, Wako-shi, Saitama 351-0198, Japan*

**Abstract**

Bitcoin is a digital currency and payment system based on classical cryptographic technologies which works without a central administrator such as in traditional currencies. It has long been questioned what the impact of quantum computing would be on Bitcoin, and cryptocurrencies in general. Here, we analyse three primary directions that quantum computers might have an impact in: mining, security, and forks. We find that in the near-term the impact of quantum computers appear to be rather small for all three directions. The impact of quantum computers would require considerably larger number of qubits and breakthroughs in quantum algorithms to reverse existing hash functions.

**Introduction**

The recent price surge of Bitcoin has captured the attention of the public about cryptocurrencies and their place in the future. Many of the attractive features of cryptocurrencies come from the lack of a third party controlling the currency, allowing for benefits such as lower transaction fees, speed, global use, and security. While its use as a day-to-day method of transaction is still rather limited, the rise of Bitcoin value has helped to attract great interest to it, which bitcoin advocates would suggest is the first step towards public acceptance. At the same time in the last few years there has been a growing wave of optimism about quantum computing. It is now not uncommon to hear comments suggesting that quantum computing technology has reached a level where scalability is within reach -- something unheard of even five years ago. It is well-known that the heart of Bitcoin is cryptography [1], and quantum computers are particularly good at difficult problems like searching and code breaking. This opens up many questions regarding how quantum computing is going to impact Bitcoin. Can a quantum computer be used to mine Bitcoins? Would quantum computers compromise the Bitcoin system? Could a quantum computer in the wrong hands be used to steal Bitcoins? This has been a concern in the Bitcoin community for some time, albeit a theoretical worry at this stage. In this paper, we break down some of these issues and see exactly what the implications of post-quantum computing world to Bitcoin are.

**Mining**

One of the technologies that Bitcoin is based on is the SHA-256, a cryptographic hashing function which turns arbitrary input data into a 256 bit string (the "hash"). This is a one-way function, so that it is easy to find the hash from an input but not the other way around. Bitcoin mining consists of the search problem of finding an input (the "nonce") combined with information of the most recent block that generates a hash that is less than a target value $T$, the maximum number that is acceptable to be considered a valid Bitcoin hash. The target value is continually being readjusted such that the average time between blocks is 10 minutes (at the time of writing the target is approximately $T = 8.9 \times 10^{11}$, much smaller than $2^{256} = 1.2 \times 10^{77}$). If it were possible to find a quantum algorithm to invert SHA-256 efficiently, then we could indeed mine Bitcoin easily. However, the value of Bitcoin comes from the difficulty of finding such solutions, which gives it "proof of work". Currently it is believed that there is no efficient algorithm, classical or quantum, which can invert SHA-256. Hence the only way is a brute force search, which classically means trying different inputs until a satisfactory solution is found.

Quantum mechanically, we have Grover search, which seems to be a perfect solution to this kind of problem, and has a quadratic quantum speedup. Let us see how well this strategy works when comparing it to mining with a classical computer. Classically, the success probability of mining a block with guesses is given by $Trt/2^{256}$, where $r$ is the hash rate (the number of guesses made per second), and $t$ is the time in seconds. For a quantum miner running Grover's algorithm the success probability is $\sin^2(2r_q t\sqrt{T/2^{256}})$ [2], where $r_q$ is the number of Grover iterations per second, which we can call the "quantum hash rate".

Now there is a different dynamic between the classical and quantum miner because Bitcoin is designed to find a new block on average every 10 minutes (=600 seconds), and hence the nature of the search problem changes in this time. In order for the Grover procedure to give a high success probability, a quantum miner should run their algorithm for a time $t$ before the problem changes, and then make a measurement. Meanwhile, the classical miner has in this time been trying as many nonces as possible. So the quantum miner is hoping that none of the classical miners have found a solution yet during the Grover evolution. Since the interval between blocks follows an exponential distribution, the probability that the block is still mineable is given by $e^{-t/600}$. Assuming a constant cost of running a quantum computer for a given amount of time, the profitability of quantum Bitcoin mining is then

$$Re^{-t/600}\sin^2(2r_q t\sqrt{T/2^{256}}) - Ct$$

where $R$ is the reward (currently equal to the price of 12.5 Bitcoins plus transaction fees) and $C$ is the cost of running the quantum computer.

Let us now estimate some plausible numbers to see whether quantum Bitcoin mining is profitable. We assume a quantum computer which costs the same to use per hour as a classical computer, and use today's Bitcoin price, block reward, and mining difficulty. We estimate that quantum Bitcoin mining becomes profitable at a quantum hash rate of 48 kilo-hashes/s. Comparing to the current best classical Bitcoin mining hardware with a hash rate of 125 kilo-hashes/s, these numbers may appear promising, but we have to keep in mind that classical Bitcoin miners can achieve enormous hash rates because the random guess mining algorithm can be quite easily parallelized. The problem is that the quantum advantage doesn't exceed the factor $\sqrt{2^{256}/T}$, no matter how many qubits one has. Thus while there is a quantum advantage, it is not insurmountable enough that classical parallelization cannot beat it. For a quantum computer with a slower hash rate than the minimally profitable 48 kilo-hashes/s, one would then need to resort to classical parallelization of quantum computers. For example, if the quantum hash rate is 3 kilo-hashes/s one would need 1300 quantum computers to be on par with classical best mining hardware that can be purchased today! Thus for quantum mining to be profitable one would need rather fast quantum hash rates, and/or a much more significant quantum speedup. This may still happen in the future, but for now classical mining seems difficult to beat.

**Security**

To ensure that Bitcoin is spent only by their rightful owners, the elliptic curve digital signature algorithm (ECDSA) is used. In short, it is based on public key cryptography where Bitcoin owners can uniquely sign transactions using their private key, and others can verify that it is genuine using their public key. Elliptic curve cryptography *is* vulnerable to quantum computing, since Shor's algorithm can be easily modified to decrypt messages sent with elliptic curves [3], i.e. a quantum computer could then be used to find the private key from a public key. This appears to expose a vulnerability, but in fact there are several safeguards built into Bitcoin which prevent this. Firstly, public keys are not revealed by your address. The Bitcoin protocol generates addresses by putting the public key through SHA-256 and then through RIPEMD-160. Since the public key is only revealed when the Bitcoins are spent, it becomes vulnerable to an attack by a quantum computer only after the public key is revealed in a transaction. This situation is readily remedied by generating a new address after each transaction (as is current best practice anyway). Once quantum computers become commonplace, it is likely most Bitcoin clients will switch to automatic key generation after each transaction. This may reduce the convenience of certain applications. For example, you will not be able to print out an address QR code and use it permanently as a cash register. Regrettably (as we will soon see) this fix is only temporary.

Another possibility is that once a public key is revealed in a pending transaction, a malicious actor, Eve, with a quantum computer could steal the bitcoins before the transaction is finalized. In principal Eve only has 10 minutes to find the private key before the transaction is finalized. In practice bitcoin transactions often sit in an unofficial pending pool (the "mem-pool") for an hour or more. Proos and Zalka estimate that for 256 bit ECDSA about 1500 qubits are required and $6 \times 10^9$ one-qubit additions are needed (Each one-qubit addition takes 9 quantum gates )[3]. Thus to execute this type of attack within an hour the quantum computer needs to perform gate operations speed of around 660 MHz. More recently Roetteler et Al finds that 2330 qubits are needed and 1.26 * 10^11 Toffoli gate operations are required (note: non-Toffoli gates are assumed to take negligible time in this work)[10]. By this estimate, despite needing more qubits, the quantum computer would only need to run at

350 MHz to pull off the attack. In either case the demands on the number of qubits and speed make this attack impossible for early generations of quantum computers.

Assuming you could break both SHA-256 and RIPEMD-160, then with an existing address it would be easy to take the money out of someone else's holdings.  However, notice that this problem is essentially the same as the problem of finding hash collisions for the purpose of adding blocks to the blockchain (i.e. mining). The theory goes that so long as the computing power needed to steal Bitcoins is much higher than the computing power needed to mine Bitcoins, anyone who could steal would instead mine. Quantum computing doesn't change this logic very much.  Doing a hash collision attack on a classical computer requires generating wallets exhaustively, whereas in a post quantum world raw inputs to the hash function can be checked and then private keys discerned after the fact. Thus a constant factor speedup in collision attacks is obtained.  Beyond that constant factor the existing security of Bitcoin against hash collision attacks still stands. Could some other vulnerability exist? While there is no proof that other vulnerabilities don't exist, currently there is no reason to believe that there is.  The only assurance we have is that Bitcoin has existed for years without anyone hacking it despite huge financial incentives to do so. While we do not rule out the possibility of novel quantum algorithms being invented for the purpose of hacking or mining, it appears highly non-trivial to say the least.

**Forks**

In a world with quantum computers the security of Bitcoin would lay entirely in the hash functions SHA-256 and RIPEMD-160 and in the slow clock cycles of early quantum computers.  There is no mathematical proof that either of these functions is not easily reversible.  Widely used hash functions have been defeated in the past (e.g. SHA-1 and MD5).  For this reason many people would see the wisdom in changing the Bitcoin public key signature system over to something that is not vulnerable to a quantum computer. Classical cryptographic primitives which are not known to be vulnerable to quantum computers do exist. Having said this, in general there is no proof that any of the post-quantum methods are actually secure against a quantum computer.  For example cryptography based on the Shortest Vector Problem (SVP) is currently considered secure.  However the SVP (like all lattice problems) has a lot of symmetry and periodicity to it. This makes it a problem that would appear well suited to quantum computing. It is conceivable that a quantum algorithm for solving SVP exists but simply hasn't been found yet. At some point there may be an arms race between the discovery of quantum algorithms and the invention of unorthodox ways of doing classical cryptography.

Bitcoin could get caught in the middle of this arms race. The result could be a series of hard forks (forward compatible protocol changes) as the Bitcoin protocol tries to keep up with the latest developments.  As we saw with the recent Bitcoin Cash fork, such changes in protocol are always controversial and dangerous for a cryptocurrency.  However, unlike contentious points of debate such block size, no one is under the illusion that ECDSA is an essential part of Bitcoin's purpose or design.  Substituting a different method should not affect scaling or decentralization. In this sense it is likely that quantum related forks will be less controversial than recent points and be accepted widely.   We note that Vitalik Buterin (inventor of Ethereum) has proposed the use of Lamport signatures, which are based on hash functions and are thus quantum resistant. In this sense making cryptocurrencies quantum-safe is an issue that is already starting to be taken seriously today.

**Conclusions**

We have found that the near-term impact of quantum computers will most probably be rather small for Bitcoin. For quantum mining, the quantum advantage is limited, hence very fast quantum hash rates would be required. For security, there is a vulnerability for pending transactions due to the use of elliptical curve cryptography, which can be broken by a variant of Shor's algorithm. However, due to time constraints during the pending transaction of approximately 10 minutes, this creates rather stringent demands on the capabilities of the quantum computer. Finally, the mere possibility that quantum computers exist could potentially destabilize Bitcoin itself through a series of forks. We speculate however that such changes to the protocol are likely to be less controversial since it is not a fundamental change in its design.

Looking into the future, one can imagine a quantum arms race between exotic forms of cryptography and quantum algorithms to break them. In such a scenario, the development of Qubitcoins -- quantum versions of Bitcoin that are based on quantum mechanics -- will become a real necessity. In a future where quantum technologies are commonplace it will make the most sense to re-invent the blockchain and transaction chain with the methods of quantum cryptography. Depending on what the future quantum computing infrastructure is like, we can have rather different scenarios for Qubitcoin. It is still not clear whether the average person will have a quantum computer or if quantum computers will exist as a cloud service and user machines remain classical. Perhaps in the future the average person will have quantum communication but full quantum computing is done on a server. As has been already explored by several works [4-7], quantum principles will change the picture significantly. Some researchers are already exploring possible qubitcoin implementations [9]. The no-cloning theorem will make it impossible to copy and distribute a decentralized ledger of qubits. The fact that qubits can't be copied or non-destructively read means they can act more like literal coins (and thus not be double-spent). One can imagine that Qubitcoin miners might support the network by doing operations which amount to quantum error correction. The use of quantum entanglement will make it possible for all participants in a network to simultaneously agree on a measurement result without a proof of work system. Until these technologies come to fruition, many scenarios can be imagined for the future of Qubitcoin. It does appear that we will need to wait for some time before we will see the Satoshi of the quantum era.

We note that after the completion of this paper we became aware of Ref. [8] containing a similar analysis.

**Acknowledgments**

This work is supported by the Shanghai Research Challenge Fund; New York University Global Seed Grants for Collaborative Research; National Natural Science Foundation of